\documentclass[aps]{revtex4}
\usepackage[dvips]{graphicx}
\usepackage{amssymb}
\begin{document}

\newcommand{\ket}[1]{|#1\rangle}
\newcommand{\bra}[1]{\langle #1|}
\newcommand{\bracket}[2]{\langle #1|#2\rangle}
\newcommand{\ketbra}[1]{|#1\rangle\langle #1|}
\newcommand{\average}[1]{\langle #1\rangle}
\newcommand{\eq}{\!=\!}
\newcommand{\maxdim}{45}
\newtheorem{theorem}{Theorem}
\newtheorem{conjecture}{Conjecture}

\title{Symmetric Informationally Complete Quantum Measurements}
\author{Joseph M.~Renes$^1$, Robin Blume-Kohout$^2$,
A.~J.~Scott$^1$, and Carlton M.~Caves$^1$\vspace{0.1in}}
\affiliation{$^1$Department of Physics and Astronomy, University of New Mexico,\\
Albuquerque, New Mexico 87131--1156, USA
\\
$^2$Theoretical Division, LANL, MS-B213, Los Alamos, New Mexico 87545, USA\vspace{0.1in}}

\date{2003 October~13}

\begin{abstract}We consider the existence in arbitrary finite
dimensions $d$ of a POVM comprised of $d^{2}$ rank-one operators all of
whose operator inner products are equal.  Such a set is called a
``symmetric, informationally complete'' POVM (SIC-POVM) and is
equivalent to a set of $d^{2}$ equiangular lines in $\mathbb{C}^d$.
SIC-POVMs are relevant for quantum state tomography, quantum
cryptography, and foundational issues in quantum mechanics. We
construct SIC-POVMs in dimensions two, three, and four.  We further
conjecture that a particular kind of group-covariant SIC-POVM exists in
arbitrary dimensions, providing numerical results up to dimension
$\maxdim$ to bolster this claim.
\end{abstract}

\maketitle

\section{Introduction}
In quantum theory, measurements are represented by \emph{positive
operator valued measures\/} (POVMs).  A POVM is termed
\emph{informationally complete\/} if its statistics determine
completely the quantum state on which the measurement is carried
out~\cite{prug77,busch91,schroek96,d'ariano03}. In order to be
maximally efficient at determining the state, such a measurement should
also be \emph{rank-one\/}; i.e., the measurement operators or
\emph{POVM elements\/} should be positive multiples of projectors onto
pure states, in which case each POVM element corresponds uniquely (up
to a phase) to a subnormalized vector in $\mathbb{C}^d$.  A
particularly appealing and potentially useful measurement is one which
is \emph{symmetric}, meaning all pairwise inner products between the
POVM elements are equal.  Such a POVM is a ``symmetric, informationally
complete positive operator-valued measure,'' or SIC-POVM for short. The
set of vectors comprising a SIC-POVM has also been studied in a very
different context, where it has a different name: it is a set of
$d^{\,2}$ equiangular lines in $\mathbb{C}^d$, first studied by Lemmens
and Seidel~\cite{lemmens73} and subsequently by many
others~\cite{Delsarte75,Koenig92,Hoggar98,koldobsky01,
Strohmer03,Koenig03,Et-Taoui00,Et-Taoui02}.  In quantum information
theory such measurements are relevant to quantum state
tomography~\cite{cfs02}, quantum cryptography~\cite{fuchssasaki03a},
and to foundational studies~\cite{fuchs02} where they would make for a
particularly interesting ``standard quantum measurement''. The
outstanding question we address in this paper is whether SIC-POVMs
exist in any finite dimension.

We conjecture that SIC-POVMs exist in all finite dimensions and,
moreover, that there exists in all finite dimensions a SIC-POVM that is
covariant under a standard representation of
$\mathbb{Z}_d\times\mathbb{Z}_d$.  To state the conjecture, let us
formalize the definition of a SIC-POVM.  The simplest definition is
that a SIC-POVM $P$ is a set of $d^{\,2}$ normalized vectors
$\ket{\phi_k}$ in $\mathbb{C}^d$ satisfying
\begin{equation}
\label{eq:sicprule}
  |\bracket{\phi_j}{\phi_k}|^2 = \frac{1}{d+1}\;,\qquad j\neq k\;.
\end{equation}
More precisely, the POVM elements of $P$ are the subnormalized
projectors $\ket{\phi_k}\bra{\phi_k}/d=\Pi_k/d$, which have pairwise
Hilbert-Schmidt inner product $(\Pi_j,\Pi_k)/d^{\,2}={\rm
Tr}[\Pi_j^\dagger\Pi_k]/d^{\,2}=1/d^{\,2}(d+1)$ for $j\ne k$.  It turns
out that the other properties of a SIC-POVM, i.e., completeness and
informational completeness, follow from Eq.~(\ref{eq:sicprule}), as we
show in section~2.

We can now state our conjecture.
\begin{conjecture}
For any dimension $d\in\mathbb{N}$, let $\{\ket{k}\}_{k=0}^{d-1}$
be an orthonormal basis for $\mathbb{C}^d$, and define
\begin{equation}
\omega=\exp(2\pi i/d)\;,\qquad
D_{\!jk}=\omega^{jk/2}\sum_{m=0}^{d-1}\omega^{jm}\ket{k\oplus m}\bra{m}\,,
\end{equation}
where $\oplus$ denotes addition modulo $d$. Then there exists a
normalized $\ket{\phi}\in\mathbb{C}^d$ such that the set
$\{D_{\!jk}\ket{\phi}\}_{j,k=1}^{d}$ is a SIC-POVM $P$.
\end{conjecture}

\noindent Analytic solutions are known for $d\eq
2,3,8$~\cite{koldobsky01}, and to this list we add $d\eq 4$.
Additionally, computer calculations reveal numerical solutions (with an
accuracy better than 1 part in $10^8$) in dimensions up to \maxdim,
some derived using the aforementioned group, but others using other
suitable groups. These results are detailed herein according to the
following plan. Section~2 states the problem in the language of frame
theory and derives a connection to the problem of finding spherical
$t$-designs. Section~3 specializes to the group-covariant case and
explains why our conjecture might be generally true. Section~4 presents
our analytic solutions for $d\eq2,3,4$, and section~5 the method of
obtaining the numerical results.  Finally, in section~6 we discuss
possible approaches to a general proof, as well as related open
questions.

\section{Frames and Spherical $t$-Designs}

The concepts of frame theory provide a simple and elegant means of
putting our problem in a general setting, for a SIC-POVM is a
particular kind of frame.  Frames are a generalization of basis sets,
with the requirements of orthogonality and normalization relaxed.  For
a finite-dimensional vector space $\mathcal{H}$, a collection of
vectors $\ket{\psi_k}\in\mathcal{H}$ is a frame if there exist
constants $0<a\leq b<\infty$ such that
\begin{equation}
  a|\bracket{\xi}{\xi}|^2\leq \sum_k|\bracket{\xi}{\psi_k}|^2\leq
  b|\bracket{\xi}{\xi}|^2
\end{equation}
for all $\ket{\xi}\in\mathcal{H}$.  Any collection of vectors is a
frame in the subspace spanned by the vectors.  The constants $a$ and
$b$ are called the \emph{frame bounds}, and if $a\eq b$, the frame is
said to be \emph{tight}. The \emph{frame operator\/} is the positive
operator
\begin{equation}
S\eq \sum_k \ket{\psi_k}\bra{\psi_k}\;.
\end{equation}
It should be immediately clear that for a tight frame $S\eq aI$.  This
tight-frame condition is equivalent to the completeness condition for
the corresponding POVM elements $\ket{\psi_k}\bra{\psi_k}/a$, and thus
rank-one POVMs and tight frames are the same mathematical object.

Now let $\mathbb{S}^d\subset\mathbb{C}^d$ be the subset consisting of
vectors that have unit norm.  Any frame can be rewritten in terms of
the corresponding normalized vectors, but tightness is not preserved
under this transformation.  For a frame
$\{\ket{\psi_k}\in\mathbb{S}^d\}_{k=1}^n$ made up of normalized
vectors, the quantity
\begin{equation}
{\rm Tr}[S^2]\eq\sum_{j,k}|\bracket{\psi_j}{\psi_k}|^2
\end{equation}
is called the \emph{frame potential}.  We consider only frames made up
of normalized vectors throughout the following.

A useful theorem due to Benedetto and Fickus~\cite{benedettofickus03}
states the following.
\begin{theorem}[Benedetto-Fickus]
\label{thm:bf} Given any $d$ and $n$, let
$\{\ket{\psi_k}\in\mathbb{S}^d\}_{k=1}^n$ be a set of normalized
vectors with frame operator S. Then
\begin{equation}
  {\rm Tr}[S^2]\geq \max(n,n^2/d\,)\;.
\end{equation}
Furthermore, the bound is achieved if and only if $\{\ket{\psi_k}\}$
consists of orthonormal vectors, when $n\le d$, or is a tight frame,
when $n\ge d$.
\end{theorem}
\noindent

{\bf Proof} The proof is so simple that we can include it here for
completeness.  Denoting the ordered eigenvalues of $S$ by
$\lambda_1\ge\lambda_2\ge\ldots\ge\lambda_d$, we first note that the
number of nonzero eigenvalues is at most $q=\min(n,d\,)$.  Thus we have
\begin{equation}
{\rm Tr}[S]=n=\sum_{k=1}^q\lambda_k
\qquad{\rm and}\qquad
{\rm Tr}[S^2]=\sum_{k=1}^q\lambda_k^2\;.
\end{equation}
Minimizing ${\rm Tr}[S^2]$ subject to the constraint ${\rm Tr}[S]=n$
gives the inequality.  Equality holds if and only if $\lambda_k=n/q$,
$k=1,\ldots,q$.  Thus for $n\le d$, $S$ is a projector onto an
$n$-dimensional subspace, implying that the vectors $\ket{\psi_k}$ are
orthogonal, and if $n\ge d$, $S=(n/d)I$, implying that the set
$\{\ket{\psi_k}\}$ is a tight frame. $\square$

\noindent
Since the frame potential of $P$ is Tr$[S^2]\eq d^{\,3}\eq n^2/d$, the
theorem establishes immediately that $P$ is a tight frame, whence $P$
is also a POVM.

For $P$ to be informationally complete, the $d^{\,2}$ operators
$\Pi_k=\ket{\phi_k}\bra{\phi_k}$ must be linearly independent so that
they span the space of operators.  The linear independence follows from
considering the rank of their Gram matrix $(\Pi_j,\Pi_k)={\rm
Tr}[\Pi_j^\dagger\Pi_k] \eq (d\delta_{jk}+1)/(d+1)$, which being
circulant (each row is a cyclic shift of the previous row), has
eigenvalues given by the Fourier transform of one of the rows. A simple
calculation reveals that due to the combination of constant term and
Kronecker delta, the eigenvalues are exactly the same as the values in
any row. Since no eigenvalues are zero, the Gram matrix has full rank,
the projection operators $\Pi_k$ are linearly independent, and $P$ is
informationally complete.

Since every rank-one POVM is a tight frame, a SIC-POVM $P$ is clearly
something more.  To fully elucidate the properties and applications of
$P$, we need to introduce \emph{spherical $t$-designs}. Building on the
result of Benedetto and Fickus, we can establish a connection between
frames and spherical $t$-designs applicable to the SIC-POVM problem.

A \emph{spherical $t$-design\/} is a set of $n$ normalized vectors
$\{\ket{\phi_k}\in\mathbb{S}^d\}$ such that the average value of any
$t$-th order polynomial $f_t(\psi)$ over the set $\{\ket{\phi_k}\}$ is
equal to the average of $f_t(\psi)$ over \emph{all\/} normalized
vectors $\ket{\psi}$. Note that if a set is a $t$-design, it is also an
$s$-design for all $s\leq t$, since an $s$-th order polynomial is also
a $t$-th order polynomial. Spherical $t$-designs were originally
developed as subsets of the real sphere $S^d$; here we apply the
concept to the set $\mathbb{S}^d$.

Let $\mathcal{H}\eq \mathbb{C}^d$, $\mathcal{H}_t$ be the $t$-fold
tensor product of such spaces, and $\mathcal{S}_t$ be the symmetric
subspace of $\mathcal{H}_t$, and consider a function
$f_t:\mathcal{H} \rightarrow\mathbb{C}$ defined as
\begin{equation}
\label{eq:polydef}
f_t(\psi)=\bra{\Psi^t} F_t \ket{\Psi^t}\;,\qquad
\ket{\Psi^t}=\ket{\psi}^{\otimes t}\;,\qquad
\ket{\psi}\in\mathcal{H}\;,
\end{equation}
where the choice of $f_t$ is equivalent to a choice of a symmetric
operator $F_t\in\mathcal{B}(\mathcal{S}_t)$. Such a function is a
$t$-th order polynomial function on $\mathcal{H}$. We can decompose
$F_t$ into a sum of product operators, i.e., $F_t\eq
\sum_k\bigotimes_{j=1}^t A_{j;k}$; thus any such function can be
decomposed into monomial terms like
\begin{equation}
\label{eq:polydecomp}
\Bigl\langle\Psi^t\Bigl|
\bigotimes_{j=1}^t A_{j}
\Bigr|\Psi^t\Bigr\rangle
=\prod_{j=1}^t\bra{\psi}A_j\ket{\psi}\;.
\end{equation}
Without loss of generality, we can restrict our attention to such
monomial functions and rewrite them as
\begin{equation}
\label{eq:polyPO} f_t(\psi)
=\prod_{j=1}^t{\rm Tr}\Bigl[A_j\ket{\psi}\bra{\psi}\Bigr]
={\rm Tr}\Biggl[\Biggl(\bigotimes_{j=1}^t A_j\Biggr)
\Pi_\psi^{\otimes t}\Biggr]\;,\qquad
\Pi_\psi=\ket\psi\bra\psi\;.
\end{equation}

Since the set $\{\ket{\phi_k}\}$ is a $t$-design if and only if
the average of any $f_t$ over $\{\ket{\phi_k}\}$ is equal to its
average over all $\ket\psi \in \mathbb{S}^d$, we are led to
compute the average of an arbitrary monomial term:
\begin{equation}
\label{eq:ave} \langle f_t\rangle
=\int\! {\rm d}\psi\,
{\rm Tr}\Biggl[\Biggl(\bigotimes_{j=1}^t A_j\Biggr)
\Pi_\psi^{\otimes t}\Biggr]
={\rm Tr}\Biggl[\Biggl(\bigotimes_{j=1}^t A_j\Biggr)
\int\!{\rm d}\psi\,\Pi_\psi^{\otimes t}\Biggr]
={\rm Tr}\Biggl[\Biggl(\bigotimes_{j=1}^t A_j\Biggr) K_t\Biggr]\;.
\end{equation}
Hence we focus on finding $K_t$, since it effectively takes the
average of $f_t$. A spherical $t$-design is then a set of vectors
for which
\begin{equation}
\label{eq:sdopdef}
S_t=\sum_{k=1}^n \ket{\Phi_k^t}\bra{\Phi_k^t}=n K_t\;,
\qquad
\ket{\Phi_k^t}=\ket{\phi_k}^{\otimes t}\;.
\end{equation}
Note that $S_t$ is the $t$-fold tensor-product analog of the frame
operator $S$.

To find the operator $K_t$, note that $K_t$ has support only on the
symmetric subspace $\mathcal{S}_t$.  Further, because $K_t$ is
invariant under any $U^{\otimes t}$ for $U\in SU(d)$, we conclude that
$K_t \propto \Pi_{\rm sym}$, the projector onto $\mathcal{S}_t$.
(Recall that $\mathcal{S}_t$ is an irreducible invariant subspace of
the group consisting of the operators $U^{\otimes t}$.)  Finally, to
determine the constant of proportionality, we consider the average of
the trivial function $f_t(\psi)\eq 1$. Equation~(\ref{eq:ave}) then
becomes Tr$[K_t]\eq 1$, and since $\mathcal{S}_t$ has dimension ${
t+d-1 \choose d-1}$, we have
\begin{equation}\label{eq:Kt}
K_t=\frac{t!(d-1)!}{(t+d-1)!}\Pi_{\rm sym}\;.
\end{equation}
For $t=1$, we see that a 1-design is a tight frame made up of
normalized vectors and, hence, also a POVM made up of equally
weighted rank-one projectors.

Equation~(\ref{eq:sdopdef}) now says that the set $\{\ket{\phi_k}\}$ is
a $t$-design if and only if the set $\{\ket{\Phi_k^t}\}$ is a tight frame
on $\mathcal{S}_t$, whence we can apply Theorem~1 to obtain the
following result.
\begin{theorem}
\label{thm:designcond}
 A set of normalized vectors $\{\ket{\phi_k}\in\mathbb{S}^d\}_{k=1}^n$
 with $n\ge{t+d-1 \choose d-1}$ forms a spherical $t$-design if and
 only if
 \begin{equation}
   {\rm Tr}[S_t^2]=
   \sum_{j,k}|\langle\phi_j|\phi_k\rangle|^{2t}
   =\frac{n^2 t!\,(d\!-\!1)!}{(t\!+\!d\!-\!1)!}\;.
 \end{equation}
 Furthermore, this value is the global minimum of ${\rm Tr}[S_t^2]$.
\end{theorem}

This theorem links the spherical $t$-design property with the
minimization of the $t$-th frame potential, Tr$[S_t^2]$.
Immediately we can infer that every SIC-POVM is a 2-design since
Tr$[S_2^2]=\sum_{j,k}|\bracket{\phi_j}{\phi_k}|^4=2d^3/(d+1)$, the
required value for a 2-design.  The converse is also true, namely,
every 2-design with $n=d^{2}$ elements is a SIC-POVM.  To show
this, let $\lambda_{jk}=|\bracket{\phi_j}{\phi_k}|^2$, $j\ne k$,
and interpret these $\lambda_{jk}$ as coordinates in
$\mathbb{R}^{d^2\!(d^2-1)}$. Using the values of the frame
potentials for a 2-design with $n=d^2$ elements, we can write
\begin{equation}
\sum_{j\neq k}\lambda_{jk}={\rm Tr}(S^2)-d^2=\frac{d^2(d^2-1)}{d+1}
\qquad{\rm and}\qquad
\sum_{j\neq k}\lambda_{jk}^2={\rm Tr}(S_2^2)-d^2=\frac{d^2(d^2-1)}{(d+1)^2}\;.
\end{equation}
The first equation describes a plane and the second a sphere. They
intersect at the single tangent point $\lambda_{jk}=1/(d+1)$, thus
showing that all 2-designs with $d^2$ elements are SIC-POVMs.
These considerations make clear that the crucial distinguishing
property of a SIC-POVM is that it is also a 2-design.  Moreover,
this ensures that minimizing the second frame potential, as we do
in the numerical work reported in section 4, yields vectors that
do indeed form a SIC-POVM.

Furthermore, $d^2$ is the smallest number of elements a 2-design can
have, so a SIC-POVM is a \emph{minimal} 2-design (see
also~\cite{koldobsky01}). Theorem~\ref{thm:designcond} does not provide
the minimum number of states, $n_{\rm min}$, for a $t$-design in $d$
dimensions, but we can establish lower bounds. For the $t=2$ case,
consider again the steps leading to the definition of the operator
$K_t$. In carrying out the average of the function $f_2$, we could have
written
\begin{equation}
\label{eq:superave} \langle f_2\rangle
={\rm Tr}\Big[A_2\!\int\!{\rm d}\psi\,
\ket{\psi}\bra{\psi}A_1\ket{\psi}\bra{\psi}\Big]
={\rm Tr}[A_2\mathcal{G}(A_1)]
\end{equation}
and thus considered the superoperator
$\mathcal{G}:\mathcal{B}(\mathbb{C}^d)\rightarrow\mathcal{B}(\mathbb{C}^d)$.
Here $\mathcal{G}(UAU^\dagger)\eq U\mathcal{G}(A)U^\dagger$ for
any $U\in SU(d)$, so by Schur's lemma, $\mathcal{G}$ is some
linear combination of projectors onto the invariant subspaces of
$U$ acting on $\mathcal{B}(\mathbb{C}^d)$.  These invariant
subspaces are (i)~the $(d^{\,2}\!-\! 1)$-dimensional subspace of
traceless operators and (ii)~the one-dimensional subspace spanned
by the identity operator $I$. Thus we can write $\mathcal{G}\eq
a\mathcal{I}+b\mathbf{I}$ where $\mathcal{I}(A)\eq A$ and
$\mathbf{I}(A)\eq {\rm Tr}[A]I$ (i.e., $\mathcal{I}$ is the
identity superoperator, and $\mathbf{I}$ projects onto the
identity operator).  To find $a$ and $b$, we first let $A_1\eq
A_2\eq I$, which gives the function $f_2(\psi)\eq 1$, so that
Eq.~(\ref{eq:superave}) yields $d(a+bd)\eq 1$. Next we consider
$A_1\eq A_2\eq \ket{\phi}\bra{\phi}$, for which $\langle
f_2\rangle\eq \int\!{\rm d}\psi\,|\bracket{\phi}{\psi}|^4=a+b$. We
can use Eqs.~(\ref{eq:ave}) and (\ref{eq:Kt}) to show that
$\langle f_2\rangle\eq 2/d(d+1)$; combined with the previous
result, this implies $a\eq b\eq 1/d(d+1)$. Therefore,
$\mathcal{G}$ has no null subspace, must be rank-$d^{\,2}$, and
cannot be constructed from less than $d^{\,2}$ linearly
independent rank-one superoperators. Similar arguments can be
applied to all spherical $t$-designs.  By similar rearrangements,
we can make several different types of operators $K_t'$, and the
rank of each serves as a lower bound on the number of vectors
required to comprise a $t$-design.

\section{Group Covariance}
\label{sec:gc}

Our results are obtained by considering group-covariant sets, so it is
appropriate to specialize to this case. The SIC-POVM $P$ is group
covariant if there exists a group $G$ with a $d$-dimensional projective
unitary representation $\{U_g\}$ such that (i)~$P$ is invariant under
any $U_g$, i.e., for any $\ket{\phi_j} \in P$ and any $U_g$,
$U_g\ket{\phi_j} \in P$ (up to a phase), and (ii)~$\{U_g\}$ acts
transitively on $P$, i.e., for any $\ket{\phi_j}, \ket{\phi_k} \in P$,
there exists $U_g$ such that $U_g\ket{\phi_j} = \ket{\phi_k}$ (also up
to a phase). Assuming group covariance simplifies the search for
SIC-POVMs.  We simply search for a fiducial vector such that
$P=\{U_g\ket{\phi}\}$ is a SIC-POVM (note that the transitivity
property implies that the order of $G$ must be at least $d^{\,2}$). To
do this, we use groups such that $\{U_g\ket{\phi}\}$ is a 1-design,
i.e., a POVM, for any normalized vector $\ket\phi$, and then we search
for a particular vector $\ket{\phi}$ such that
$|\bra{\phi}U_g\ket{\phi}|^2\eq 1/(d+1)$ for all $g\neq e$.  All other
inner products are then guaranteed to have this value due to the group
action.

We suspect the case of group covariance to be general for the following
reason. Consider the map $\alpha:\mathbb{S}^{d}\rightarrow
\mathcal{B}(\mathbb{C}^d)$ that takes a normalized vector to the
corresponding projector, i.e., $\alpha(\ket{\phi_j})\eq
\ket{\phi_j}\bra{\phi_j}$.  Now consider the operators
\begin{equation}
\sigma_j\eq\sqrt{{d\over d-1}}
\left(\ket{\phi_j}\bra{\phi_j}-{I\over d}\right)\;.
\end{equation}
Being both traceless and Hermitian, these operators lie in a
subspace of $\mathcal{B}(\mathbb{C}^d)$ that is isomorphic to
$\mathbb{R}^{d^{2}\!-1}$; indeed, since $(\sigma_j,\sigma_j)=1$,
they all lie on the unit sphere in $\mathbb{R}^{d^{\,2}-1}$.  This
sphere is a generalization of the Bloch sphere for two-dimensional
systems, the difference being that for $d>2$, not all operators on
the sphere are images of vectors in $\mathbb{S}^d$ under the map
$\alpha$.  From the SIC-POVM condition~(\ref{eq:sicprule}), one
finds immediately that
\begin{equation}
(\sigma_j,\sigma_k)=-\frac{1}{d^{\,2}\!-\!1}\qquad\forall\,j\neq k\;.
\end{equation}
This is the condition for the $d^{\,2}$ operators $\{\sigma_j\}$ to
form a regular simplex in $\mathbb{R}^{d^{\,2}-1}$, whose automorphism
group is the permutation group $S_{d^2}$. Given this result, some group
covariance seems natural. One is tempted to think that from here it is
a simple matter to establish the existence of the set $P$. This is not
the case, however, as working in the operator space obscures the very
difficult task of determining when a given operator is the image of
some element of $\mathbb{S}^d$ under the map $\alpha$. In the same
vein, most of the elements of the permutation group cannot be
represented in this framework as unitary transformations of
$\mathbb{C}^d$; thus, while we know that any $G$ satisfying the
conditions above must be a subgroup of $S_{d^{2}}$, it is not obvious
which subgroups are candidates.

The outstanding choice for $G$ is the group
$\mathbb{Z}_d\times\mathbb{Z}_d$, as described in the conjecture. We
note in passing that using this group to find a SIC-POVM makes $P$ a
Gabor or Weyl-Heisenberg frame, this being the definition of such a
frame \cite{hanlarson00}; such a SIC-POVM $P$ is useful in defining
finite-dimensional analogs of the familiar $P$ and $Q$
quasidistributions of infinite dimensions. The group's usefulness here
stems from the fact that, for any normalized
$\ket{\psi}\in\mathbb{S}^d$,
\begin{equation}
\label{eq:grouptfcond}
  S_\psi=\sum_{jk}D_{\!jk}\ket{\psi}\bra{\psi}D_{\!jk}^\dagger=dI\;,
\end{equation}
a fact readily checked by direct calculation.

The property in Eq.~(\ref{eq:grouptfcond}) of producing a 1-design for
any input state is quite general. The following argument is adapted
from Proposition~3 of~\cite{werner01}. Any set of $d^{\,2}$ orthogonal
unitary operators $T_j$, thus satisfying Tr$[T_j^\dagger T_k]\eq
d\delta_{jk}$, is a complete set for expanding operators in
$\mathcal{B}(\mathbb{C}^d)$; the unitary operators $\{D_{jk}\}$ are one
example of operators that satisfy this orthogonality condition.  It is
a simple matter to turn the completeness relation into $\sum_k T_k C
T_k^\dagger\eq d\,{\rm Tr}[C]I$ for any operator $C$.  Simply consider
the inner product of two arbitrary operators $A$ and $B$. The
completeness relation means that
\begin{equation}
  (A,B)=\frac{1}{d}\sum_k(A,T_k)(T_k,B)\;.
\end{equation}
Setting $A\eq \ket{\phi_1}\bra{\phi_2}$ and $B\eq
\ket{\psi_1}\bra{\psi_2}$ (which we can do without loss of generality
because such outer products span $\mathcal{B}(\mathbb{C}^d))$, we find
\begin{equation}
  \bracket{\phi_1}{\psi_1}\bracket{\psi_2}{\phi_2}=
  \frac{1}{d}\sum_k\bra{\phi_1}T_k\ket{\phi_2}\bra{\psi_2}T_k^\dagger\ket{\psi_1}\;,
\end{equation}
from which it follows that $\sum_k
T_k\ket{\phi_2}\bra{\psi_2}T_k^\dagger=d\,\bracket{\psi_2}{\phi_2}$,
whence the result follows.

Thus the property of producing a 1-design regardless of the fiducial
$\ket{\phi_0}$ is common to all groups of size $d^{\,2}$ whose
representation operators are a complete, orthogonal set.  Such groups
were introduced by Knill in connection with quantum error-correcting
codes and are called ``nice error bases'' or unitary error
bases~\cite{knill96a, knill96b}. Klappenecker and R\"{o}tteler have
kindly detailed all such nice error bases up to dimension 10, so we can
apply them to the problem at
hand~\cite{klappiroetteler02,klappiroetteler03,nebonline}.  Only the
nice error bases associated with the group
$\mathbb{Z}_d\times\mathbb{Z}_d$ exist in every dimension, thus
accounting for our focus on this group.


\section{Analytic SIC-POVMs}
Here we concentrate specifically on using the group
$\mathbb{Z}_d\times\mathbb{Z}_d$.  Fixing the representation operators
$D_{jk}$ of this group we can determine the set of fiducial vectors
that under the group action make a SIC-POVM.  From this we can
determine also the number of distinct SIC-POVMs generated by our fixed
representation.  In three dimensions there are an uncountably infinite
number of such covariant SIC-POVMs, but in two dimensions there are
just two, and in four, 16.  We write the fiducial state as
\begin{equation}
  \ket{\phi}=\sum_k r_k e^{i\theta_k}\ket{k}\;,
\end{equation}
where we can, of course, immediately choose $\theta_0=0$.

\subsection{$d=2$}

The two solutions, represented as column vectors in the standard basis, are
\begin{equation}
\left\{
\frac{1}{\sqrt{6}}\left(
\begin{array}{c}
\sqrt{3+\sqrt{3}}
\vspace{.05in}\\
e^{i\pi/4}\sqrt{3-\sqrt{3}}
\end{array}
\right),
\frac{1}{\sqrt{6}}
\left(
\begin{array}{c}
-\sqrt{3-\sqrt{3}}
\vspace{.05in}\\
e^{i\pi/4}\sqrt{3+\sqrt{3}}
\end{array}
\right)
\right\}\;.
\end{equation}
These have a simple interpretation on the Bloch sphere, where the
nontrivial group operators are simply rotations by $\pi$ about the
$x$, $y$, and $z$ axes, respectively. Then the Bloch vectors of
the two fiducial states are $\pm(1,1,1)/\sqrt{3}$, and the two
SIC-POVM states thus formed are regular tetrahedra, each one
related to the other by inversion of the Bloch vectors.

\subsection{$d\eq 3$}

For $r_0$ satisfying $1/\sqrt2<r_0<\sqrt{2/3}$, define
\begin{equation}
r_\pm(r_0) = \frac{1}{2}r_0\pm \frac{1}{2}\sqrt{2-3r_0^2}\;.
\end{equation}
Hence $0<r_-\leq 1/\sqrt{6}\leq r_+ <1/\sqrt{2}<r_0\leq\sqrt{2/3}$. The
complete set of fiducial states, represented as column vectors in the
standard basis, is then
\begin{eqnarray}
&\mbox{}&\left\{
\left(
\begin{array}{c}
r_0 \\ r_+ e^{i\theta_1} \\ r_- e^{i\theta_2}
\end{array}
\right),
\left(
\begin{array}{l}
\text{plus all vectors formed}\\ \text{by permuting of elements}
\end{array}
\right)
\Bigg|\,
\theta_1,\theta_2\in
\bigg\{\frac{\pi}{3},\pi,\frac{5\pi}{3}\bigg\},
\frac{1}{\sqrt{2}}<r_0\leq\sqrt{\frac{2}{3}}
\right\}
\nonumber \\
&\mbox{}&\phantom{xxxxxxxxxx}
\bigcup
\left\{
\left(
\begin{array}{c}
1/\sqrt{2} \\ e^{i\theta_1}/\sqrt{2} \\ 0
\end{array}
\right),
\left(
\begin{array}{l}
\text{plus all vectors formed}\\ \text{by permuting of elements}
\end{array}
\right)
\Bigg|\,0\leq\theta_1<2\pi\right\}\;.
\end{eqnarray}

\subsection{$d=4$}

Now let
\begin{equation}
r_0 = \frac{1-1/\sqrt{5}}{2\sqrt{2-\sqrt{2}}}\;,
\qquad
r_1=(\sqrt{2}-1)r_0\;,
\qquad
r_\pm = \frac{1}{2}\sqrt{1+1/\sqrt{5}\pm\sqrt{1/5+1/\sqrt{5}}}\;,
\end{equation}
along with
\begin{equation}
a = \arccos\frac{2}{\sqrt{5+\sqrt{5}}}\;,
\qquad
b = \arcsin\frac{2}{\sqrt{5}}\;,
\end{equation}
and define the set
\begin{eqnarray}
\Omega&\equiv&
\Bigg\{\bigg((-1)^m(a/2+b/4)+\pi(m+2n+7j+1)/4\,,\,\pi(2k+1)/2\, ,\,
\nonumber\\
&\mbox{}&\phantom{xx}
(-1)^m(-a/2+b/4)+\pi(m+2n+3j+4k+1)/4\bigg)\Bigg|\,
j,k,m=0,1 \text{ and } n=0,\dots,3\Bigg\}\;.
\end{eqnarray}
The complete set of fiducial states, represented as column vectors in
the standard basis, is now
\begin{eqnarray}
\left\{
\left(
\begin{array}{c}
r_0 \\ r_+e^{i\theta_+} \\ r_1e^{i\theta_1} \\ r_-e^{i\theta_-}
\end{array}
\right),
\left(
\begin{array}{c}
r_0 \\ r_-e^{i\theta_-} \\ r_1e^{i\theta_1} \\ r_+e^{i\theta_+}
\end{array}
\right),
\left(
\begin{array}{l}
\text{plus all vectors formed}\\ \text{by cycling of elements}
\end{array}
\right)
\Biggm|\,
(\theta_+,\theta_1,\theta_-) \in \Omega\right\}\;.
\end{eqnarray}

\section{Numerical SIC-POVMs}

Because analytic solutions to the SIC-POVM condition
(\ref{eq:sicprule}) are so few, our conjectures are based almost
entirely on numerical evidence (even the $d=4$ solution was originally
inspired by close examination of numerical solutions). To find
numerical instances of $P$, we simply minimize the second frame
potential Tr$[S_2^2]$ over sets of $d^{\,2}$ normalized vectors
generated by a representation of $\mathbb{Z}_d\times\mathbb{Z}_d$ from
a vector $\ket{\phi}$.  It is also possible to vary independently the
$d^{\,2}$ elements of $P$, but this is much less efficient; taking
advantage of the group-covariance conjecture permits us to search a
space of $O(d)$ complex parameters instead of $O(d^{\,3})$ complex
parameters.

The quantity that we minimize, $\sum_{j,k}{\left|\bra\phi
D_{\!jk}\ket\phi\right|^4}$, is proportional to the frame potential
because of the group covariance. Since it is a quartic function of
$\ket\phi$, we have to use numerical methods to minimize it, using
either Mathematica (simpler) or C++ (much faster).  The method used is
an adaptive conjugate gradient method; this has the advantage of
converging with exponential rapidity to a local minimum, but the
disadvantage of being insensitive to global conditions.  As a result,
the most time-intensive portion of the computation by far is
identifying one of the global minima among the many local minima.

Once the correct minimum is located, we quickly obtain $P$ such
that Eq.~(\ref{eq:sicprule}) is satisfied to an accuracy of
$10^{-8}$.  The sole exception to this rule is $d=3$ (where an
exact analytic solution is known): in $d=3$ there exists a
continuously infinite family of solutions, and this degeneracy
makes numerical solution difficult. For every dimension between
$d=5$ and $d=\maxdim$, however, we have found
$\mathbb{Z}_d\times\mathbb{Z}_d$-covariant solutions to within
machine precision~\cite{sicponline}.

Additionally, in small dimensions, one can attempt an exhaustive search
for \emph{all\/} possible $\mathbb{Z}_d\times\mathbb{Z}_d$-covariant
SIC-POVMs, by simply running the minimization many times with differing
presumptive fiducial states, tabulating all the while the distinct
SIC-POVM fiducial states found. Table 1 lists the results for the
number of distinct SIC-POVMs.
\begin{table}[h]
\label{table:sicpsets}
\begin{tabular}{c|c}
$d$ & \hspace{.05in}\#(SIC-POVMs)\\\hline
2 & 2\\
3 & $\infty$\\
4 & 16\\
5 & 80\\
6 & 96\\
7 & 336\\
\end{tabular}
\caption{Number of SIC-POVM sets generated by a fixed
representation of the group $\mathbb{Z}_d\times\mathbb{Z}_d$ in
dimensions two through seven. The infinity in dimension three is
uncountable.}
\end{table}

Finally, we have tested some of the other nice error bases tabulated by
Klappenecker and R\"{o}tteler.  These are also easy handled, and
although not all groups were tested, at least four groups were found to
generate SIC-POVM sets.  In the notation of the library of small groups
used by GAP3, GAP4, and MAGMA, these groups are G(36,11), G(36,14),
G(64,8), and G(81,9).  Each of these solutions has an accuracy of
$10^{-15}$ in the individual vector inner products.  Perhaps
surprisingly, many of the tabulated groups do not seem to yield
group-covariant SIC-POVMs.

\section{Odds and Ends}
A rigorous proof of existence of SIC-POVMs in all finite dimensions
seems tantalizingly close, yet remains somehow distant.  Although the
numerical evidence makes very clear the relevance of the group
$\mathbb{Z}_d\times\mathbb{Z}_d$, this is not definitively established.
Given the apparent importance of $\mathbb{Z}_d\times\mathbb{Z}_d$, it
would seem to be just a short step to some general form for an operator
whose eigenvectors could be a fiducial state, but a proof by this
method has not been forthcoming.  For instance, in three dimensions the
Fourier transform operator has an eigenvector that is a fiducial state
(the one associated with the eigenvalue $i$), but this does not hold in
general.  In five dimensions a fiducial vector can be found among the
degenerate eigenvectors of a particular $\mathbb{Z}_3$ subgroup of the
normalizer of $\mathbb{Z}_d\times\mathbb{Z}_d$ in $SU(d)$, but there is
no such subgroup at all in the normalizer for dimension seven.  The
group-theoretic structure of SIC-POVMs is exceedingly rich, however,
and ongoing efforts to understand the full automorphism group of a
SIC-POVM might yield insights into operators that yield fiducial
states.  We hope that by establishing the framework and providing
motivating numerical results in this paper, a proof might be completed.

Finally, regardless of whether their existence can be proved
rigorously, SIC-POVMs appear to exist in many dimensions.  They might
be of use in many areas of quantum information theory---signal
ensembles, quasidistributions on discrete phase space, error-correcting
codes, and quantum measurement, just to name a few.  In addition, they
appear to be connected to a number of interesting mathematical
problems, including spherical codes and the putative existence of
mutually unbiased bases.

\vspace{.25in} \noindent \textbf{Acknowledgements} The authors
gratefully acknowledge assistance and advice from H.~Barnum, C.~A.
Fuchs, K.~Manne, J.~P. Paz, W.~K. Wootters, and W.~H. Zurek. This
work was supported in part by Office of Naval Research Grant
No.~N00014-00-1-0578.


\begin{thebibliography}{99}

\bibitem{prug77}
E.~Prugove\u{c}ki, International Journal of Theoretical Phyics,
{\bf 16}, 321 (1977).

\bibitem{busch91}
P.~Busch, International Journal of Theoretical Physics, {\bf 30},
1217 (1991).

\bibitem{schroek96}
F.~E.~Schroek, \emph{Quantum Mechanics on Phase Space}, (Kluwer,
Dordrecht, the Netherlands, 1996).

\bibitem{lemmens73}
P.~W.~H. Lemmens and J.~J. Seidel, ``Equiangular Lines,'' Journal
of Algebra {\bf 24}, 494--512 (1973).

\bibitem{Delsarte75}
P.~Delsarte, J.~M. Goethels, and J.~J. Seidel, ``Bounds for
Systems of Lines and Jacobi Polynomials,'' Philips Research
Reports {\bf 30}, 91--105 (1975).

\bibitem{Koenig92}
H.~K\"onig and N.~Tomczak-Jaegermann, ``Norms of Minimal
Projections,'' {\tt arXiv:math.FA/9211211}.

\bibitem{Hoggar98}
S.~G. Hoggar, ``64 Lines from a Quaternionic Polytope,''
Geometriae Dedicata {\bf 69}, 287--289.

\bibitem{koldobsky01}
A.~Koldobsky and H.~K\"onig, ``Aspects of the Isometric Theory of
Banach Spaces,'' in {\sl Handbook of the Geometry of Banach
Spaces}, Vol.~1, edited by W.~B. Johnson and J.~Lindenstrauss,
(North Holland, Dordrecht, 2001), pp.~899--939.

\bibitem{Strohmer03}
T.~Strohmer and R.~Heath, ``Grassmanian Frames with Applications
to Coding and Communication,''  Appl.\ Comp.\ Harm.\ Anal {\bf
14}, 257-275 (2003).

\bibitem{Koenig03}
H.~K\"onig, ``Cubature Formulas on Spheres,'' available online at
{\tt
http://analysis.math.uni-kiel.de/\discretionary{}{}{}koenig/\discretionary{}{}{}preprints.html}.

\bibitem{Et-Taoui00}
B.~Et-Taoui, ``Equiangular Lines in $C^r$,'' Indagationes
Mathematicae {\bf 11}, 201--207 (2000).

\bibitem{Et-Taoui02}
B.~Et-Taoui, ``Equiangular Lines in $C^r$ (Part II),''
Indagationes Mathematicae {\bf 13}, 483--486 (2002).

\bibitem{cfs02}
C.~M. Caves, C.~A. Fuchs, and R.~Schack, ``Unknown Quantum States:\
The Quantum de Finetti Representation,'' J. Math.\ Phys.\ {\bf
43}, 4537--4559 (2002).

\bibitem{fuchssasaki03a}
C.~A. Fuchs and M.~Sasaki, ``Squeezing Quantum Information through
a Classical Channel:\ Measuring the `Quantumness' of a Set of
Quantum States,'' Quant.\ Info.\ Comp.\ {\bf 3}, 377--404 (2003).

\bibitem{fuchs02}
C.~A. Fuchs, ``Quantum Mechanics as Quantum Information (and only
a little more),'' {\tt quant-ph/0205039}, and private
communication.

\bibitem{d'ariano03}
G.~M. D'Ariano, P.~Perinotti, and M.~F. Sacchi, ``Informationally
Complete Measurements and Groups Representation,'' {\tt
arXiv:quant-ph/0310013}.

\bibitem{werner01}
R.~F.~Werner, ``All Teleportation and Dense Coding Schemes,''
Journal of Physics A {\bf 34}, 7081--7094 (2001).

\bibitem{benedettofickus03}
J.~J.~Benedetto and M.~Fickus, Advances in Computational
Mathematics, {\bf 18}, 357--385 (2003).

\bibitem{klappiroetteler02}
A.~Klappenecker and M.~R\"{o}tteler. ``Beyond Stabilizer Codes I:
Nice Error Bases,'' IEEE Transactions on Information Theory
{\bf IT-48}, 2392--2395 (2002).

\bibitem{klappiroetteler03}
A.~Klappenecker and M.~R\"{o}tteler. `` Unitary error bases:
Constructions, equivalence, and applications,'' Applied Algebra,
Algebraic Algorithms and Error-Correcting Codes, Proceedings; {\bf
2643}, 139--149 (2003).

\bibitem{nebonline}
Klappenecker and R\"{o}tteler maintain an online catalog of nice
error bases available at
\texttt{http://\discretionary{}{}{}faculty.cs.tamu.edu/\discretionary{}{}{}klappi/ueb/ueb.html}.

\bibitem{knill96a}
E.~Knill, ``Non-binary Unitary Error Bases and Quantum Codes,'' LANL
report LAUR-96-2717, available at {\tt arXiv:quant-ph/9608048}.

\bibitem{knill96b}
E.~Knill, ``Group Representations, Error Bases and Quantum Codes,''
LANL report LAUR-96-2807, available at {\tt arXiv:quant-ph/9608049}.

\bibitem{hanlarson00}
D.~Han and D.~R.~Larson, ``Frames, Bases, and Group
Representations,'' Memoirs of the American Mathematical Society
{\bf 147}, 697 (2000).

\bibitem{sicponline}
Available online at {\tt
http://info.phys.unm.edu/papers/reports/sicpovm.html}

\end{thebibliography}
\end{document}